\documentclass{INTERSPEECH2023}
\usepackage{subcaption}
\usepackage{enumitem}
\usepackage{commath}

\interspeechcameraready 

\title{Unsupervised speech intelligibility assessment with utterance level alignment distance between teacher and learner Wav2Vec-2.0 representations}
\name{Nayan Anand, Meenakshi Sirigiraju, Chiranjeevi Yarra}
\address{Speech Lab, Language Technologies Research Center (LTRC), IIIT, Hyderabad, 500032, India}

\email{nayan.anand@research.iiit.ac.in, meenakshi.sirigiraju@research.iiit.ac.in, chiranjeevi.yarra@iiit.ac.in}

\begin{document}

\maketitle
 
\begin{abstract}



Speech intelligibility is crucial in language learning for effective communication. Thus, to develop computer-assisted language learning systems, automatic speech intelligibility detection (SID) is necessary. Most of the works have assessed the intelligibility in a supervised manner considering manual annotations, which requires cost and time; hence scalability is limited. To overcome these, this work proposes an unsupervised approach for SID. The proposed approach considers alignment distance computed with dynamic-time warping (DTW) between teacher and learner representation sequence as a measure to separate intelligible versus non-intelligible speech. We obtain the feature sequence using current state-of-the-art self-supervised representations from Wav2Vec-2.0. We found the detection accuracies as 90.37\%, 92.57\% and 96.58\%, respectively, with three alignment distance measures -- mean absolute error, mean squared error and cosine distance (equal to one minus cosine similarity). 

\end{abstract}


\section{Introduction}

The number of language learners worldwide has been rising consistently in recent times \cite{duolingo2022}, fueled by factors like globalization, education, career prospects, cultural curiosity, and personal growth. Speech intelligibility is vital for language learners as it allows them to effectively communicate their thoughts, ideas and needs. However, several factors, such as pronunciation, speech disorders, prosody, and grammar \cite{de2002intelligibility, derwing1997accent}, can affect the intelligibility of their speech. Manual assessment of intelligibility involves listening to the speech and deciding scores based on the ratio of correctly identified words to the total number of words spoken \cite{flipsen2006measuring}. It can be detected at the sentence or discourse level as well. But subjective intelligibility tests are costly and laborious. Hence the need for an automatic assessment of speech intelligibility is important which is reliable and cost-effective.

Supervised learning approaches based on labeled data enable the detection of speech intelligibility \cite{ spille2018predicting}. These methods involve training machine learning models using labels given by experts for learners' speech data, allowing them to accurately classify and assess the intelligibility of speech. In \cite{jia2020deep,dong2020attention,zezario2020stoi}, the deep learning (DL) models are used as regression functions to predict objective evaluation scores using Short-Time Objective Intelligibility (STOI), and Speech Transmission Index (STI). And \cite{chen2021inqss,andersen2018nonintrusive,pedersen2020neural} to predict human subjective ratings from human listening tests. 
In \cite{zezario2022mti}, a model MTI-Net is proposed using convolutional bidirectional long short-term memory (CNN-BLSTM) with an attention mechanism. It is trained
using a multitask learning criterion and predicts the Intelligible score as one of its target outputs. While these works in speech intelligibility assessment rely on ratings from experts, our approach leverages an unsupervised technique.



Our work is motivated by the goal of employing an unsupervised approach to speech intelligibility detection. To accomplish this, we utilize the alignment distance computed using dynamic-time warping (DTW) between the Wav2Vec-2.0 self-supervised representations of the teacher and learner. This alignment distance serves as a metric to differentiate between intelligible and non-intelligible speech. In our experiments, we explore three different distance measures: Mean absolute error (MAE), Mean squared error (MSE), and Cosine distance (CD). We compared the performance of these measures with the baseline methods: Majority Class Voting (MCV), and Random Selection (RS).



\section{Dataset}
VoisTUTOR corpus \cite{yarra2019voistutor} was used for the experiments performed in this work. This dataset contains English speech recordings from teacher and 16 learners (8 males, 8 females) for 1676 unique stimuli. The stimuli ranges from single-word utterances to multiple words forming simple, compound, and complex sentences. The selection of stimuli was done from spoken English materials. The chosen stimuli cover various aspects of pronunciation, and include phonological elements such as fricatives, stops, nasals, glides \& laterals, consonant sequences, vowels, diphthongs, and semi-vowels. The learners belonged to six different Indian native languages: Kannada, Telugu, Tamil, Malayalam, Hindi, and Gujarati. Each audio recording in the dataset was then annotated as either intelligible (1) or not (0). The annotations were obtained from a spoken English teacher with 25 years of teaching experience in professional English speaking skills in India. It is found that a percentage of 88.08\% are marked as intelligible and the remaining are non-intelligible.  In addition to the annotations, in this data, the teacher recordings were also obtained from the same teacher. We consider all the speech samples from the teacher, which are clear and understandable.
\section{Methodology}

The proposed approach identifies intelligible and non-intelligible speech in an unsupervised manner considering the alignment distance between teacher and learner. The alignment distance is computed using DTW on representation sequences obtained from the Wav2Vec-2.0 model for teacher and learner audios. For the distance computation, we consider three types of distance measures, namely, 1) MAE, 2) MSE, and 3) CD. With this approach, we hypothesize that the distance between audios of the same stimuli is closer compared to that between audios of different stimuli. Thus, we believe the proposed approach could be useful for speech intelligibility detection task. This section summarizes the above components as follows. In sub-section \ref{Wav2vec2.0:3.1}, we describe the Wav2Vec-2.0 model and the motivation for using representation from the model. In sub-section \ref{utt_lvl_alignment_compute}, the three types of distance measures are discussed. In sub-section \ref{DTW_section}, we describe the utterance level alignment distance computation.
\subsection{Wav2Vec-2.0}
\label{Wav2vec2.0:3.1}
Wav2Vec-2.0 \cite{baevski2020wav2vec_cite_27} is a state-of-the-art model for obtaining representation sequence for an input raw audio considering a self-supervised representation learning framework. These representations have been considered in many end-to-end speech recognition tasks \cite{yi2020applying_cite_2,jain2023wav2vec2_cite_3}.
Self-supervised representation learning approach in Wav2Vec-2.0 takes unstructured raw audio data and learns the representations that could disentangle the phonemic aspects in the audio. These representations are then used for fine-tuning the downstream tasks including speech recognition \cite{fan2020exploring_cite_16}.
Wav2Vec-2.0 focuses on capturing complex patterns from waveform; introducing non-linearity by choosing activation functions such as GELU \cite{hendrycks2016gaussian_cite_23}. This in turn enhances its generalizability and ensures a robust representation of the waveform.

 Due to this effective learning process, Wav2Vec-2.0 repesentations have been considered in a wide variety of tasks besides speech recognition such as Speaker recognition \cite{vaessen2022fine_cite_1} Speaker adaptation \cite{baskar2022speaker_cite_5}, Speaker verification \cite{fan2020exploring_cite_16}, Cross-lingual knowledge transfer \cite{yi2021transfer_cite_4}, Mispronunciation detection \cite{yang2022improving_cite_7,xu2021explore_cite_18,peng2021study_cite_19}, Voice activity detection \cite{kunevsova2023multitask_cite_9}, Prosodic boundary detection \cite{kunevsova2022detection_cite_11}, Emotion identification  \cite{wang2021fine_cite_12,pepino2021emotion_cite_14} as well as Non-verbal vocalization detection \cite{tzirakis2023large_cite_13}. Furthermore, it has been widely explored for medical domain tasks such as Stuttering \cite{sheikh2022introducing_cite_8,grosz2022wav2vec2_cite_10, bayerl2022detecting_cite_28}, Alzheimer detection \cite{gauder2021alzheimer_cite_17} as well as developing system for rating children speech with speech sound disorder \cite{getman2022wav2vec2_cite_6}.

We obtain Wav2Vec-2.0 representations for teacher and learner speech utterances. Since Wav2Vec-2.0 captures linguistic features  \cite{fan2020exploring_cite_16}, in terms of discriminative properties between similar and dissimilar phonemes, we hypothesize that these representations can be used for speech intelligibility detection task.

\subsection{Distance measures}
\label{utt_lvl_alignment_compute}
This work explores the following distance measures to act as a cost function for computing utterance level alignments between teacher-learner speech pairs for given stimuli: \\
\vspace{-0.2cm}

\textbf{Mean absolute error (MAE):} It is a metric used to quantify the average absolute difference between any two vectors $t_x$ and $l_y$ of dimension N. It ranges between [0,$\infty$).
\vspace{-0.3cm}
\begin{equation}
\mathrm{MAE(t_x,l_y)}=\frac{1}{N} \sum_{i=1}^N\left|t_{x_i}-l_{y_i}\right|
\end{equation} \\

\vspace{-0.6cm}
\textbf{Mean squared error (MSE):} It is a metric used to quantify the average squared difference between any two vectors $t_x$ and $l_y$ of dimension N. It is highly prone to outliers as compared to MAE due to the squaring of errors. It ranges between [0,$\infty$).
\vspace{-0.3cm}
\begin{equation}
\mathrm{MSE(t_x,l_y)}=\frac{1}{N} \sum_{i=1}^N\left(t_{x_i}-l_{y_i}\right)^2
\end{equation} \\
\vspace{-0.5cm}
\vspace{-0.4cm}
\textbf{Cosine Distance (CD):} It is a metric that quantifies the angular dissimilarity between any two vectors $t_x$ and $l_y$ of dimension N. \cite{senoussaoui2013study}. It remains unaffected by the vector magnitudes and has a range of [0,2].

\vspace{-0.5cm}
\begin{equation}
\begin{split}
\mathrm{CD(t_x,l_y)}= 
1-\dfrac{\sum_{i=1}^N t_{x_i} l_{y_i}}{\sqrt{\sum_{i=1}^N (t_{x_i})^2} \sqrt{\sum_{i=1}^N (l_{y_i})^2}}
\end{split}
\end{equation}

The aforementioned cost functions are capable to measure either the angular dissimilarity or differences between two vectors, which aligns with our main objective of assessing the dissimilarity between the teacher and learner speech for speech intelligibility detection.

\subsection{Utterance level alignment distance computation}
\label{DTW_section}
For a given stimuli, the number of frames for the teacher and learner's speech is different. Hence, we cannot obtain a direct one-to-one mapping between their frames in a sequential manner as it would always leave some frames unmatched. Furthermore, there is always the possibility of different phones being stretched temporally for a teacher-learner speech pair. This gives rise to an uneven distribution of frames across phonemes. To address this issue, the dynamic programming-driven DTW \cite{sakoe1978dynamic_cite_30} algorithm is considered; which is highly time efficient with a focus on cost optimization. It obtains the best possible frame level alignments between each of the teacher-learner speech pairs (T, L) varied across all stimuli. 

Assuming a given teachers' and learners' speech representation have X and Y frames i.e ($T$= [ $t_1$,$t_2$...$t_X$ ] and $L$= 
 [ $l_1$,$l_2$...$l_Y$ ]). All possible combinations of teacher and learner frames from  $T$ and $L$ are passed to  Equation 4. sequentially in the form of ($t_x$, $l_y$ ). At a given time step,  $t_x$ and $l_y$ correspond to a single frame from $T$ and $L$ respectively.
 They are passed as input to the distance measure `c' of Equation 4 where `c' can be chosen from Equation 1, Equation 2, or Equation 3.
In Equation 4, C(x,y) corresponds to the accumulated alignment cost, and c($t_x$, $l_y$) represents the local alignment cost. Here x and y correspond to frame indexes which map ($t_x$, $l_y$) to ($T$, $L$). The optimal utterance level alignment is thus obtained by minimizing the  local and accumulated alignment cost at each time step.

\vspace{-0.6cm}
\begin{equation}
\label{equation_for_dtw}
\vspace{-0.15cm}
\begin{array}{r}
C(x, y)=c\left({t}_x, {l}_y\right)+\min \left[C(x-1, y),\right. \\
\left.C(x, y-1), C(x-1, y-1)\right]
\end{array}
\end{equation}

\begin{figure}[!htbp]
\centering

\includegraphics[trim=2cm 2cm 4cm 2cm, width=.40\textwidth  ]{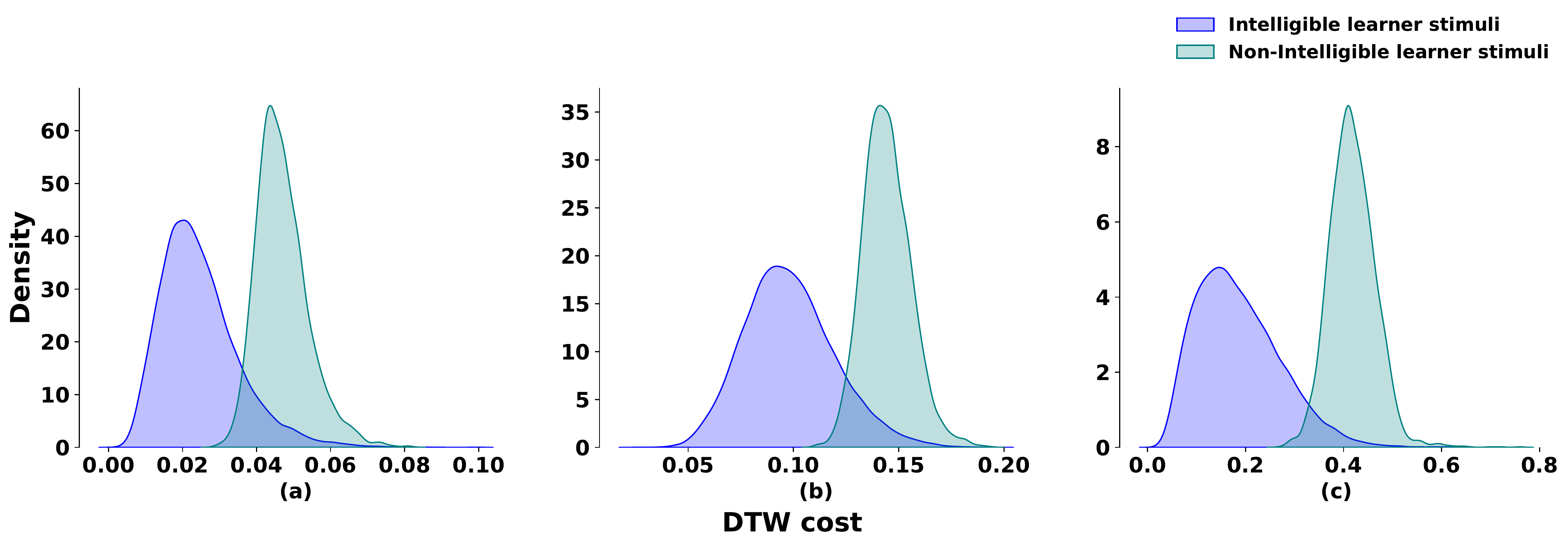}

\caption{ Distributions from utterance level alignment score obtained for the cost functions (a) Mean Absolute Error(MAE), (b) Mean Squared error (MSE) and (c) Cosine Distance (CD) } 
\label{fig1z}
\end{figure}

Figure 1 shows the distribution of alignment distance obtained for intelligible and non-intelligible speech classes on the entire learner's data with the three distance measures. From the figure, it is observed that the distance is higher for non-intelligible compared to intelligible speech for all three measures. Also, the overlap of the distance values is minimal between the two speech classes. Thus, we believe that the proposed utterance alignment distance could be useful for speech intelligibility detection.

\section{Experiments}

We perform the experiments on the entire learners' data in the VoisTUTOR dataset. Out of which 5\% of the data is used for threshold ($\tau$) computation using EER (Equation \ref{equn6}) criterion. The remaining 95\% of the data is considered as a test set for evaluating the performance of the proposed approach. We consider classification accuracy as the performance measure using ground-truth class labels available in the data. We perform the classification considering all three distance measures separately i.e. for each distance we compute $\tau$ and used it for the classification of the respective alignment distance values. We compute the alignment distances considering teacher audio in the VoisTUTOR dataset in a stimuli-specific manner separately with all three distance measures.

\vspace{-0.3cm}
\begin{equation}
\label{equn4}
F A R=\frac{\text { Number of false acceptances }}{\text { Number of identification attempts }}
\end{equation}
\vspace{-0.6cm}

\begin{equation}
\label{equn5}
F R R=\frac{\text { Number of false rejections }}{\text { Number of identification attempts }}
\end{equation}

\vspace{-0.2cm}


\begin{equation}
\label{equn6}
E E R=\frac{F A R+F R R}{2}
\end{equation}

\textbf{Baselines:}
We consider the following baselines for comparing the performance of the proposed approach for all three distance measures.

\emph{Majority Class Voting (MCV):}  
We assign all the learner's speech in the test set as an intelligible class since the majority class is intelligible. With this assignment, we compute the classification accuracy and used it for the comparison.

\emph{Random Selection (RS):} In this, we assign the labels randomly to the learners' speech samples considering the binary distribution. The parameter for the binary distribution is chosen from the class distribution using the randomly selected 5\% subset of the VoisTUTOR dataset. With these randomly assigned labels, we compute the classification accuracy and compare it with the results.



\begin{table}[!htbp]
\centering

\caption{Classification Accuracy percentage between intelligible and non-intelligible speech obtained across various comparison methods (CD: Cosine Distance, MAE: Mean Absolute Error, MSE: Mean Squared Error, MCV: Majority Class Voting, RS: Random Selection) }
\small
\begin{tabular}{|l|l|l|l|l|}
\hline
\multicolumn{1}{|p{1.1cm}|}{\textbf{CD}} & \multicolumn{1}{p{1.2cm}|}{\textbf{MAE}} & \multicolumn{1}{p{1.2cm}|}{\textbf{MSE}} & \multicolumn{1}{p{1.1cm}|}{\textbf{MCV}} & \multicolumn{1}{p{1.1cm}|}{\textbf{RS}} \\ \hline
96.58          & 90.37             & 92.57           & 88.08        & 85.44                        \\ \hline
\end{tabular}

\end{table}
\vspace{-0.4cm}

\begin{table}[!htbp]
\centering
\caption{Classification accuracy percentage between intelligible and non-intelligible speech obtained across various comparison methods with a focus on different phoneme categories (CD: Cosine Distance, MAE: Mean Absolute Error, MSE: Mean Squared Error, MCV: Majority Class Voting, RS: Random Selection) }
\small
\resizebox{\linewidth}{!}{
\begin{tabular}{|l|c|c|c|c|c|}
\hline
\textbf{Phoneme Category}   & \textbf{CD} & \textbf{MAE} & \textbf{MSE} & \textbf{MCV} & \textbf{RS} \\ \hline
\textbf{Fricatives}  & 95.90 & 91.16 & 91.74 & 83.56 & 80.88 \\ \hline
\textbf{Stops}       & 96.21 & 89.63 & 90.43 & 87.23 & 84.29 \\ \hline
\textbf{Nasals}      & 96.93 & 91.01 & 91.01 & 90.98 & 87.95 \\ \hline
\textbf{Semi-Vowels} & 96.26 & 90.59 & 92.73 & 90.23 & 87.05 \\ \hline
\textbf{Glides}      & 95.94 & 89.66 & 90.63 & 85.04 & 81.30 \\ \hline
\textbf{Vowels}      & 92.87 & 90.27 & 92.80 & 79.51 & 75.83 \\ \hline
\textbf{Diphthongs}  & 96.23 & 94.86 & 94.55 & 81.46 & 78.21 \\ \hline
\textbf{Consonant Clusters} & 93.00       & 84.70        & 89.58        & 82.01        & 78.15       \\ \hline
\end{tabular}
}
\end{table}
\vspace{0.6 cm}

\section{Results}

 The following results for different classification approaches were obtained:
\vspace{0.5 cm}
 
\begin{figure}[!htbp]

\vspace{-0.90 cm}
\captionsetup{skip=0pt}  
\centering
\vspace{-0.9 cm}
\label{flowchart10}
\includegraphics[trim=4cm -1cm 4cm -6cm, width=.32\textwidth ,height=.55\textheight]
{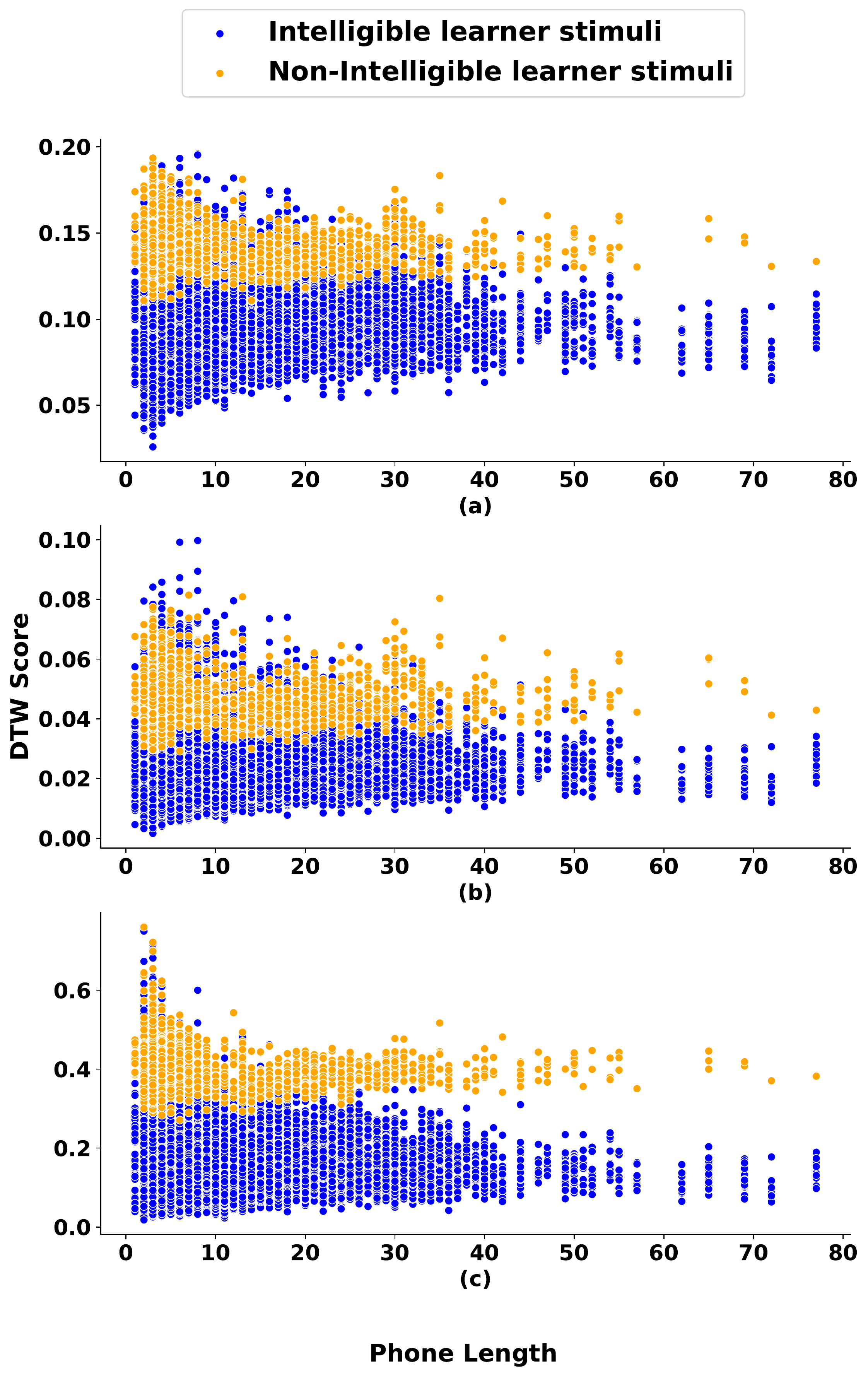}

\caption{(Utterance level alignment distance for intelligible and non-intelligible learner stimuli across phone length using distance measures (a) Mean Absolute Error (MAE), (b)
Mean Squared error (MSE) and (c) Cosine Distance (CD) }
\vspace{-0.4 cm}

\end{figure}


\subsection{Overall performance}
  Table 1 shows the classification accuracies obtained with the proposed approach using all three distance measures -- CD, MAE, and MSE along with MCV and RS baselines. From the table, it is observed that the accuracies obtained with the proposed approach with all three distance measures are higher than that with the two baselines. The highest relative improvement is found to be 9.65\% and 13.03\% respectively compared with MCV and RS baselines. This indicates the benefit of the proposed approach for speech intelligibility detection task. Among the three distance measures considered, it is observed that the classification is best obtained using CD with a classification accuracy of 96.58\% followed by MSE having a classification accuracy of 92.57\%, and lastly MAE with a classification accuracy of 90.37\%. This could be due to the normalization involved in the metric computation. Thus, it limits the alignment distance span to a limited range. 

\subsection{Phoneme category-specific performance}
 Table 2 shows the classification accuracies obtained across different phoneme categories with the proposed approach with all three distance measures -- CD, MAE, and MSE along with MCV and RS baselines. The best classification accuracies across categories are as follows: 96.93\%  using the CD for Nasals, 94.86\% using MAE, and 94.55\% using MSE for Diphthongs respectively. The highest relative improvement with baseline MCV is found to be 18.13\% using CD and 16.45\% using MAE for Diphthongs, and 16.71\% using MSE for Vowels respectively. When compared to the RS baseline, the highest relative improvement is found to be 23.04\% using CD, 21.30\% using MAE for Diphthongs, and 22.37\% using MSE for Vowels. 


 Diphthongs and vowels are one of the most critical phonemes categories for determining the intelligibility of a speech utterance.
This is because, with even a single vowel or diphthong replacement in speech, a drastic shift in the meaning of the utterance can be observed.
Furthermore, even a slight deviation in pronunciation across these two categories may lead to perceptual discrimination of the utterance. 
Hence we observed a relatively higher classification accuracy across these categories when the alignment distances are limited to a smaller range.



\subsection{Analysis with phone specific length}  

Figure 2 illustrates the separation between intelligible and non-intelligible speech samples by calculating the alignment distance at the utterance level for stimuli with varying phone lengths of learner and teacher speech. This distinction across different phone lengths indicates that the proposed approach is capable of generalization, as it remains effective regardless of the number of phones. The sub-figures 2(a), 2(b), and 2(c) illustrate the separation between intelligible and non-intelligible speech with varying phone lengths using MSE, MAE, and CD respectively. The X-axis in these sub-plots mark phone length whereas Y-axis represents the utterance level alignment distance between learner and teacher feature sequences. Blue dots indicate intelligible learner speech samples whereas orange dots represent non-intelligible learner speech samples. 
The separation boundary between the two classes is found to
be consistent across all phone lengths for all three distance measures. These sub-figures complement the observations from Table 1 with the best separation being obtained for CD followed by MSE and MAE.



\subsection{Analysis with illustrative examples}

We hypothesize the intersection of phone boundaries for teacher's and intelligible learner's speech  to lie on the DTW path thereby ensuring that the best possible alignment has been achieved. However, for non-intelligible learner speech, this might not be true as it may be better aligned with some other speech stimuli.  
To verify this hypothesis the DTW paths between teacher and learner feature sequences are plotted for intelligible as well as non-intelligible learner speech for a given stimuli. The intersection of the teacher's and the learner's corresponding  phone boundaries are then traced. This can be visualized in Figure 3 where the black line represents DTW path between teacher and learner whereas green and purple dotted lines represent phone boundaries of teacher and learner respectively. The intersection of phone boundaries for teacher and learner has been highlighted by red circles.


\begin{figure}[!htbp]
\centering
\includegraphics[trim=2.5cm 2cm 4cm 0cm, width=.48\textwidth  ]{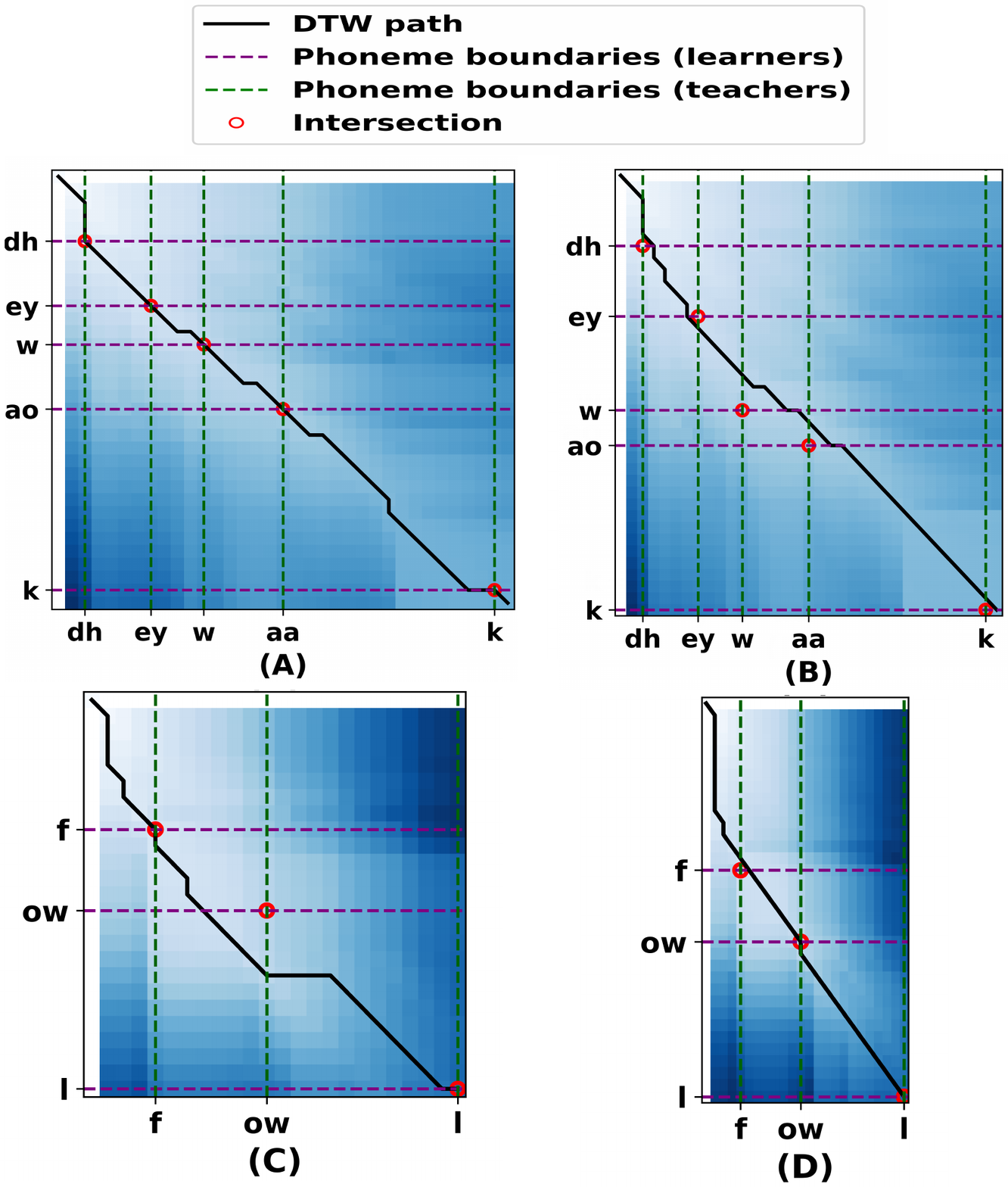}

\vspace{0.5 cm}
\caption{DTW between teacher \& learner feature sequences for 
(a) Actual intelligible classified as intelligible, 
(b) Actual non-intelligible  classified as non-intelligible, (c) Actual intelligible classified as non-intelligible, and 
(d) Actual non-intelligible classified as intelligible}
\vspace{-0.7 cm}
\label{fig:block_dia2}
 \end{figure}

The following four scenarios were explored while visualizing the DTW path:
\begin{enumerate}
    \item When the learner's speech with intelligible ground truth label is classified as intelligible 
    \item When the learner's speech with a non-intelligible ground truth label is classified as non-intelligible
    \item When the learner's speech with intelligible ground truth label is classified as non-intelligible 
    \item When the learner's speech with a non-intelligible ground truth label is classified as intelligible
\end{enumerate}

 The outcomes of these can be visualized in Figure 3  where sub-figures (3(a) \& 3(b)) illustrate the successful detection scenario whereas  sub-figures (3(c) \& 3(d)) showcase the failed case scenarios. It is observed that when learner speech with intelligible ground truth labels is classified as intelligible, the intersection of phoneme boundaries for teacher and learner for any uttered sequence falls on the DTW path. In all other cases, the intersection of phoneme boundaries for teacher and learner is observed to deviate from the DTW path.

\section{Conclusion }


In this work, we propose an unsupervised approach for speech intelligibility detection task. The alignment distance computed using DTW between Wav2Vec-2.0 representations of teacher and learner speech is used as a measure to distinguish between intelligible and non-intelligible speech. The experimental results demonstrate that the proposed approach outperforms the baseline methods MCV and RS. 
Among the three distance measures used (MSE, MAE, and CD), the CD gives the best results for experiments performed on the overall dataset as well as with data subsets having a focus on different phoneme categories. 
Notably, the proposed approach is found to be generalizable across stimuli with varying phoneme lengths. We performed our experiments on learners' speech from six different nativities of the VoisTUTOR corpus. In the future, we aim to extend this work to more nativities.


\section{Acknowledgements}

We would like to thank IHub and IIIT Hyderabad for their continued help and support toward the fulfillment of this work. 
\if
\else

\fi

\bibliographystyle{IEEEtran}
\bibliography{mybib}

\begin{thebibliography}{10}
\providecommand{\url}[1]{#1}
\csname url@samestyle\endcsname
\providecommand{\newblock}{\relax}
\providecommand{\bibinfo}[2]{#2}
\providecommand{\BIBentrySTDinterwordspacing}{\spaceskip=0pt\relax}
\providecommand{\BIBentryALTinterwordstretchfactor}{4}
\providecommand{\BIBentryALTinterwordspacing}{\spaceskip=\fontdimen2\font plus
\BIBentryALTinterwordstretchfactor\fontdimen3\font minus
  \fontdimen4\font\relax}
\providecommand{\BIBforeignlanguage}[2]{{%
\expandafter\ifx\csname l@#1\endcsname\relax
\typeout{** WARNING: IEEEtran.bst: No hyphenation pattern has been}%
\typeout{** loaded for the language `#1'. Using the pattern for}%
\typeout{** the default language instead.}%
\else
\language=\csname l@#1\endcsname
\fi
#2}}
\providecommand{\BIBdecl}{\relax}
\BIBdecl

\bibitem{duolingo2022}
\BIBentryALTinterwordspacing
C.~Blanco, ``Duolingo language report,'' 2022. [Online]. Available:
  \url{https://blog.duolingo.com/2022-duolingo-language-report/}
\BIBentrySTDinterwordspacing

\bibitem{de2002intelligibility}
M.~S. De~Bodt, M.~E. H.-D. Huici, and P.~H. Van De~Heyning, ``Intelligibility
  as a linear combination of dimensions in dysarthric speech,'' \emph{Journal
  of communication disorders}, vol.~35, no.~3, pp. 283--292, 2002.

\bibitem{derwing1997accent}
T.~M. Derwing and M.~J. Munro, ``Accent, intelligibility, and
  comprehensibility: Evidence from four l1s,'' \emph{Studies in second language
  acquisition}, vol.~19, no.~1, pp. 1--16, 1997.

\bibitem{flipsen2006measuring}
P.~Flipsen~Jr, ``Measuring the intelligibility of conversational speech in
  children,'' \emph{Clinical linguistics \& phonetics}, vol.~20, no.~4, pp.
  303--312, 2006.

\bibitem{spille2018predicting}
C.~Spille, S.~D. Ewert, B.~Kollmeier, and B.~T. Meyer, ``Predicting speech
  intelligibility with deep neural networks,'' \emph{Computer Speech \&
  Language}, vol.~48, pp. 51--66, 2018.

\bibitem{jia2020deep}
X.~Jia and D.~Li, ``A deep learning-based time-domain approach for
  non-intrusive speech quality assessment,'' in \emph{2020 Asia-Pacific Signal
  and Information Processing Association Annual Summit and Conference (APSIPA
  ASC)}.\hskip 1em plus 0.5em minus 0.4em\relax IEEE, 2020, pp. 477--481.

\bibitem{dong2020attention}
X.~Dong and D.~S. Williamson, ``An attention enhanced multi-task model for
  objective speech assessment in real-world environments,'' in \emph{ICASSP
  2020-2020 IEEE International Conference on Acoustics, Speech and Signal
  Processing (ICASSP)}.\hskip 1em plus 0.5em minus 0.4em\relax IEEE, 2020, pp.
  911--915.

\bibitem{zezario2020stoi}
R.~E. Zezario, S.-W. Fu, C.-S. Fuh, Y.~Tsao, and H.-M. Wang, ``Stoi-net: A deep
  learning based non-intrusive speech intelligibility assessment model,'' in
  \emph{2020 Asia-Pacific Signal and Information Processing Association Annual
  Summit and Conference (APSIPA ASC)}.\hskip 1em plus 0.5em minus 0.4em\relax
  IEEE, 2020, pp. 482--486.

\bibitem{chen2021inqss}
Y.-W. Chen and Y.~Tsao, ``Inqss: a speech intelligibility assessment model
  using a multi-task learning network,'' \emph{arXiv preprint
  arXiv:2111.02585}, 2021.

\bibitem{andersen2018nonintrusive}
A.~H. Andersen, J.~M. De~Haan, Z.-H. Tan, and J.~Jensen, ``Nonintrusive speech
  intelligibility prediction using convolutional neural networks,''
  \emph{IEEE/ACM Transactions on Audio, Speech, and Language Processing},
  vol.~26, no.~10, pp. 1925--1939, 2018.

\bibitem{pedersen2020neural}
M.~B. Pedersen, A.~H. Andersen, S.~H. Jensen, and J.~Jensen, ``A neural network
  for monaural intrusive speech intelligibility prediction,'' in \emph{ICASSP
  2020-2020 IEEE International Conference on Acoustics, Speech and Signal
  Processing (ICASSP)}.\hskip 1em plus 0.5em minus 0.4em\relax IEEE, 2020, pp.
  336--340.

\bibitem{zezario2022mti}
R.~E. Zezario, S.-w. Fu, F.~Chen, C.-S. Fuh, H.-M. Wang, and Y.~Tsao,
  ``Mti-net: A multi-target speech intelligibility prediction model,''
  \emph{arXiv preprint arXiv:2204.03310}, 2022.

\bibitem{yarra2019voistutor}
C.~Yarra, A.~Srinivasan, C.~Srinivasa, R.~Aggarwal, and P.~K. Ghosh,
  ``voistutor corpus: A speech corpus of indian l2 english learners for
  pronunciation assessment,'' in \emph{2019 22nd Conference of the Oriental
  COCOSDA International Committee for the Co-ordination and Standardisation of
  Speech Databases and Assessment Techniques (O-COCOSDA)}.\hskip 1em plus 0.5em
  minus 0.4em\relax IEEE, 2019, pp. 1--6.

\bibitem{baevski2020wav2vec_cite_27}
A.~Baevski, Y.~Zhou, A.~Mohamed, and M.~Auli, ``wav2vec 2.0: A framework for
  self-supervised learning of speech representations,'' \emph{Advances in
  neural information processing systems}, vol.~33, pp. 12\,449--12\,460, 2020.

\bibitem{yi2020applying_cite_2}
C.~Yi, J.~Wang, N.~Cheng, S.~Zhou, and B.~Xu, ``Applying wav2vec2. 0 to speech
  recognition in various low-resource languages,'' \emph{arXiv preprint
  arXiv:2012.12121}, 2020.

\bibitem{jain2023wav2vec2_cite_3}
R.~Jain, A.~Barcovschi, M.~Yiwere, D.~Bigioi, P.~Corcoran, and H.~Cucu, ``A
  wav2vec2-based experimental study on self-supervised learning methods to
  improve child speech recognition.'' \emph{IEEE Access}, 2023.

\bibitem{fan2020exploring_cite_16}
Z.~Fan, M.~Li, S.~Zhou, and B.~Xu, ``Exploring wav2vec 2.0 on speaker
  verification and language identification,'' \emph{arXiv preprint
  arXiv:2012.06185}, 2020.

\bibitem{hendrycks2016gaussian_cite_23}
D.~Hendrycks and K.~Gimpel, ``Gaussian error linear units (gelus),''
  \emph{arXiv preprint arXiv:1606.08415}, 2016.

\bibitem{vaessen2022fine_cite_1}
N.~Vaessen and D.~A. Van~Leeuwen, ``Fine-tuning wav2vec2 for speaker
  recognition,'' in \emph{ICASSP 2022-2022 IEEE International Conference on
  Acoustics, Speech and Signal Processing (ICASSP)}.\hskip 1em plus 0.5em minus
  0.4em\relax IEEE, 2022, pp. 7967--7971.

\bibitem{baskar2022speaker_cite_5}
M.~K. Baskar, T.~Herzig, D.~Nguyen, M.~Diez, T.~Polzehl, L.~Burget,
  J.~{\v{C}}ernock{\`y} \emph{et~al.}, ``Speaker adaptation for wav2vec2 based
  dysarthric asr,'' \emph{arXiv preprint arXiv:2204.00770}, 2022.

\bibitem{yi2021transfer_cite_4}
C.~Yi, J.~Wang, N.~Cheng, S.~Zhou, and B.~Xu, ``Transfer ability of monolingual
  wav2vec2. 0 for low-resource speech recognition,'' in \emph{2021
  International Joint Conference on Neural Networks (IJCNN)}.\hskip 1em plus
  0.5em minus 0.4em\relax IEEE, 2021, pp. 1--6.

\bibitem{yang2022improving_cite_7}
M.~Yang, K.~Hirschi, S.~D. Looney, O.~Kang, and J.~H. Hansen, ``Improving
  mispronunciation detection with wav2vec2-based momentum pseudo-labeling for
  accentedness and intelligibility assessment,'' \emph{arXiv preprint
  arXiv:2203.15937}, 2022.

\bibitem{xu2021explore_cite_18}
X.~Xu, Y.~Kang, S.~Cao, B.~Lin, and L.~Ma, ``Explore wav2vec 2.0 for
  mispronunciation detection.'' in \emph{Interspeech}, 2021, pp. 4428--4432.

\bibitem{peng2021study_cite_19}
L.~Peng, K.~Fu, B.~Lin, D.~Ke, and J.~Zhang, ``A study on fine-tuning wav2vec2.
  0 model for the task of mispronunciation detection and diagnosis.'' in
  \emph{Interspeech}, 2021, pp. 4448--4452.

\bibitem{kunevsova2023multitask_cite_9}
M.~Kune{\v{s}}ov{\'a} and Z.~Zaj{\'\i}c, ``Multitask detection of speaker
  changes, overlapping speech and voice activity using wav2vec 2.0,'' in
  \emph{ICASSP 2023-2023 IEEE International Conference on Acoustics, Speech and
  Signal Processing (ICASSP)}.\hskip 1em plus 0.5em minus 0.4em\relax IEEE,
  2023, pp. 1--5.

\bibitem{kunevsova2022detection_cite_11}
M.~Kune{\v{s}}ov{\'a} and M.~{\v{R}}ez{\'a}{\v{c}}kov{\'a}, ``Detection of
  prosodic boundaries in speech using wav2vec 2.0,'' in \emph{Text, Speech, and
  Dialogue: 25th International Conference, TSD 2022, Brno, Czech Republic,
  September 6--9, 2022, Proceedings}.\hskip 1em plus 0.5em minus 0.4em\relax
  Springer, 2022, pp. 377--388.

\bibitem{wang2021fine_cite_12}
Y.~Wang, A.~Boumadane, and A.~Heba, ``A fine-tuned wav2vec 2.0/hubert benchmark
  for speech emotion recognition, speaker verification and spoken language
  understanding,'' \emph{arXiv preprint arXiv:2111.02735}, 2021.

\bibitem{pepino2021emotion_cite_14}
L.~Pepino, P.~Riera, and L.~Ferrer, ``Emotion recognition from speech using
  wav2vec 2.0 embeddings,'' \emph{arXiv preprint arXiv:2104.03502}, 2021.

\bibitem{tzirakis2023large_cite_13}
P.~Tzirakis, A.~Baird, J.~Brooks, C.~Gagne, L.~Kim, M.~Opara, C.~Gregory,
  J.~Metrick, G.~Boseck, V.~Tiruvadi \emph{et~al.}, ``Large-scale nonverbal
  vocalization detection using transformers,'' in \emph{ICASSP 2023-2023 IEEE
  International Conference on Acoustics, Speech and Signal Processing
  (ICASSP)}.\hskip 1em plus 0.5em minus 0.4em\relax IEEE, 2023, pp. 1--5.

\bibitem{sheikh2022introducing_cite_8}
S.~Sheikh, M.~Sahidullah, F.~Hirsch, and S.~Ouni, ``Introducing ecapa-tdnn and
  wav2vec2. 0 embeddings to stuttering detection,'' in \emph{Submitted to
  Interspeech 2022}, 2022.

\bibitem{grosz2022wav2vec2_cite_10}
T.~Gr{\'o}sz, D.~Porjazovski, Y.~Getman, S.~Kadiri, and M.~Kurimo,
  ``Wav2vec2-based paralinguistic systems to recognise vocalised emotions and
  stuttering,'' in \emph{Proceedings of the 30th ACM International Conference
  on Multimedia}, 2022, pp. 7026--7029.

\bibitem{bayerl2022detecting_cite_28}
S.~P. Bayerl, D.~Wagner, E.~N{\"o}th, and K.~Riedhammer, ``Detecting
  dysfluencies in stuttering therapy using wav2vec 2.0,'' \emph{arXiv preprint
  arXiv:2204.03417}, 2022.

\bibitem{gauder2021alzheimer_cite_17}
L.~Gauder, L.~Pepino, L.~Ferrer, and P.~Riera, ``Alzheimer disease recognition
  using speech-based embeddings from pre-trained models.'' in
  \emph{Interspeech}, 2021, pp. 3795--3799.

\bibitem{getman2022wav2vec2_cite_6}
Y.~Getman, R.~Al-Ghezi, K.~Voskoboinik, T.~Gr{\'o}sz, M.~Kurimo, G.~Salvi,
  T.~Svendsen, and S.~Str{\"o}mbergsson, ``Wav2vec2-based speech rating system
  for children with speech sound disorder,'' in \emph{Interspeech}, 2022.

\bibitem{senoussaoui2013study}
M.~Senoussaoui, P.~Kenny, T.~Stafylakis, and P.~Dumouchel, ``A study of the
  cosine distance-based mean shift for telephone speech diarization,''
  \emph{IEEE/ACM Transactions on Audio, Speech, and Language Processing},
  vol.~22, no.~1, pp. 217--227, 2013.

\bibitem{sakoe1978dynamic_cite_30}
H.~Sakoe and S.~Chiba, ``Dynamic programming algorithm optimization for spoken
  word recognition,'' \emph{IEEE transactions on acoustics, speech, and signal
  processing}, vol.~26, no.~1, pp. 43--49, 1978.

\end{thebibliography}

\end{document}